\documentclass[a4paper,twocolumn,english,aps,prl,showpacs]{revtex4-1}
\usepackage[T1]{fontenc}
\usepackage[latin1]{inputenc}
\usepackage{graphicx,color}
\usepackage{amsmath}
\makeatletter

\usepackage{babel}

\begin{document}

\title{A Model for the Dissipative Conductance in Fractional  Quantum Hall States}

\author{N. d'Ambrumenil$^1$, B.I. Halperin$^2$ 
and R.H. Morf$^3$}

\affiliation{$^1$Physics Department, University of Warwick, Coventry CV4
7AL, United Kingdom \\ 
$^2$Physics Department, Harvard University, Cambridge, Massachusetts 02138, USA\\
$^3$Paul Scherrer Institute, CH-5232 Villigen, Switzerland
}

\date{\today}

\begin{abstract}
We present a model of dissipative transport in the fractional quantum Hall regime. Our 
model takes account of tunneling through saddle points in the effective
potential for excitations created by impurities. 
We predict the temperature range over which activated
behavior is observed and explain why this range nearly
always corresponds to around a factor two in temperature
in both integer quantum Hall and fractional quantum Hall
systems.  We identify the ratio of the gap observed in the activated 
behavior and thetemperature of the inflection point in the Arrhenius plot as an important diagnostic
for determining the importance of tunneling in real samples.
\end{abstract}
\maketitle

The energy gap and an incompressible ground state are essential
components of all quantum Hall systems \cite{Laughlin81,Laughlin83}. 
Estimates of this gap have usually been found by fitting the
temperature dependence of the longitudinal resistance 
of the highest mobility samples to the standard Arrhenius form 
\cite{Willett88,Du93,Dean09}. 
However, progress towards an
understanding of how the measured activation gap 
relates to the intrinsic gap has
been held up by the lack of a detailed microscopic model of 
the effect of disorder on the dissipative transport.

The Arrhenius form, $\sigma = \sigma_0 e^{-\Delta/2T}$, is found
only over a small range in temperature  
and the estimates of the gap
are lower than theoretical expectations \cite{Willett88,Du93,MorfdA03,Dean08}.
Where it has been possible to extract estimates, $\sigma_0$ has
generally been lower by 
around a factor of 2 than the predicted 
$2(qe)^2/h$ per square \cite{Polyakov95}, where $qe$ is the charge
carried by the quasiparticles (QPs) or quasiholes (QHs) \cite{Clarketal88,Katayama94}.

The gap $\Delta$  extracted from experiments has often been interpreted as
a zero-temperature mobility gap. Specifically, this equates $\Delta$ to
the energy necessary to create a pair of separated oppositely charged
quasiparticles in presumably well-defined extended single-quasiparticle
states, which can then carry charge across the system \cite{Clarketal90,Dean09,wan05,Murthy09}.  Here,
however, we advance a different picture of thermally activated transport
in a quantized Hall state in a modulation doped sample.  Our approach 
generalizes the model of \cite{Polyakov95} to
include the effect of thermally assisted tunneling and to account for 
the compressible screening regions \cite{Efros88_2,Pikus-Efros94} which
are nucleated when the gap is smaller than the unscreened potential due
to ionized donors. 
The conductance measurements are controlled by the 
saddle-point gap energy, $\Delta_s$, which is the energy required to
excite excitations at a saddle point in this potential. 
The presence of screening regions means that
$\Delta_s$ will normally be significantly less than the gap in ideal homogeneous incompressible
systems, $\Delta_h$, studied using exact diagonalizations, by an amount which is not the result of a simple
broadening of levels, $\Gamma$ \cite{Clarketal90,Du93}, and with no simple connection to the zero
field mobility \cite{Dean08}.

Our model explains why activated behavior is normally only seen over a small
regime in temperature with a factor of around two between the highest and
lowest temperatures and predicts that the prefactor, $\sigma_0$,
of the Arrhenius form is reduced from its ``ideal'' value of $2 (qe)^2/h$.
We show that the gap, $\Delta_i$, obtained from the Arrhenius plots,
and the temperature range (in units of $\Delta_i$),
over which activated behavior is observed, can be used to estimate 
$\Delta_s$.  We find $\Delta_s/\Delta_h \approx 0.55$ for the 
strongest lowest Landau level (LL) states of \cite{Du93} and
$\Delta_s/\Delta_h \approx 0.4$, for the strongest state at $\nu=5/2$ of \cite{Dean09}.

The samples studied experimentally typically
have the donor ions in layer(s) set back from the 2DEG.
The  setback distance, $d$, sets 
the length scale for the potential 
fluctuations in the electron gas created by the ionized donors. 
In most samples and at the filling fractions
and densities involved, $d \gtrsim  \ell_q=\ell_0/\sqrt{q}$. Here
$\ell_q$ is the effective magnetic length of the QP of QH, and $\ell_0 = \sqrt{\hbar/eB}$.
Modeling of the positioning of the ions in the donor layers
suggests that  $\Delta_h$ is smaller than the unscreened impurity potential, 
and predicts that  the system breaks up into 
regions of incompressible fluid separating 
compressible regions  containing  excitations,
which partially screen the impurity potential \cite{Efros92,Pikus-Efros94}.
The quantized Hall conductance of the system 
is then the response of the percolating regions of incompressible 
fluid.

When $ \ell_q \ll d$, 
and in the absence of screening regions,
the dissipative response near the 
center of the quantum Hall plateau is due
to the transfer of thermal excitations across saddle points 
in the energy landscape set up by the ionized donors \cite{Polyakov95}.
Each saddle point acts as a resistor in two separate resistor networks.  
In one, a saddle connects together the p-type regions (QH-rich)
and in the other it connects the n-type (QP-rich) regions.
The conductance in each case is given by $(qe)^2 h^{-1} \exp{(-E_s^{h,p}/kT)}$, where
$E_s^{h}$ and $E_s^{p}$ are the heights of the saddle point for QHs
and QPs with $E_s^{h} + E_s^{p} = \Delta_s$.
If the fluctuations of the impurity potential are symmetrically
distributed about the mean,  Dykhne's theorem \cite{Dykhne71} gives
that the logarithm of the conductance
of each network is the average of the logarithm of the conductance at
each node and gives an overall response of $2 (qe)^2 h^{-1} \exp{(-\Delta_s/2kT})$.

\begin{figure}[t]
\includegraphics[width=2.8in]{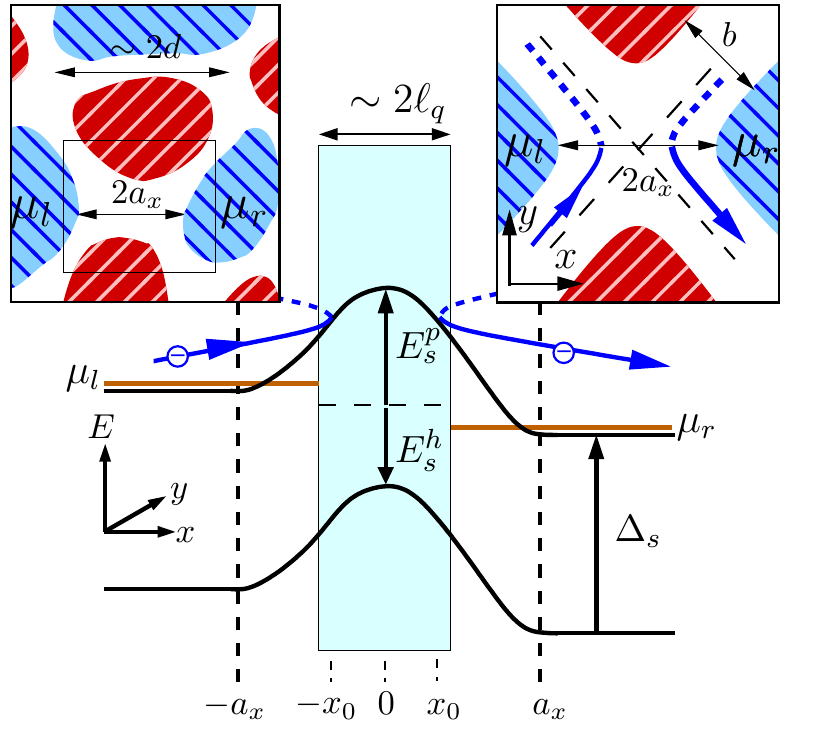}
\caption{\label{fig:bup} 
Band alignment and particle flow across a typical saddle point.
Upper left: Break-up of a quantum Hall system.
Incompressible fluid at filling fraction $\nu$ (white background)
separates 
compressible regions in which
QPs (dark diagonals on lighter shading) and QHs (light diagonals on
darker shading) are nucleated 
\cite{Efros92}. 
Upper Right: Tunnelling of
a QP through the saddle point from
an equipotential line about 
the left region (chemical potential $\mu_l$) to one about the right.
Main figure: Energy band alignment.
For QP excitations with energies $E<E_s^p$, 
the transmission probability is significant if the point of closest
approach to the saddle point, $x_0$, lies within a
zone of width $\sim 2 \ell_q$ about the saddle point. 
Potential variations in the incompressible region 
exceeding $\Delta_s$  are not possible. They would simply nucleate
carriers and reduce the size of the incompressible region.
}

\end{figure}

If $\ell_q$ becomes comparable to $d$ or $a_x$, where
$2a_x$ is the width of the incompressible region at 
a saddle (see Fig~\ref{fig:bup}), the main dissipative process still 
involves carriers crossing saddle
points, but now there
may be significant tunneling across the saddle point. 
There is then a range of temperatures in which the response appears activated, 
with an activation energy reduced from $E_s$ by an amount which depends on $a_x/\ell_q$
but not on temperature.

\begin{figure}[t]
\includegraphics[width=3in]{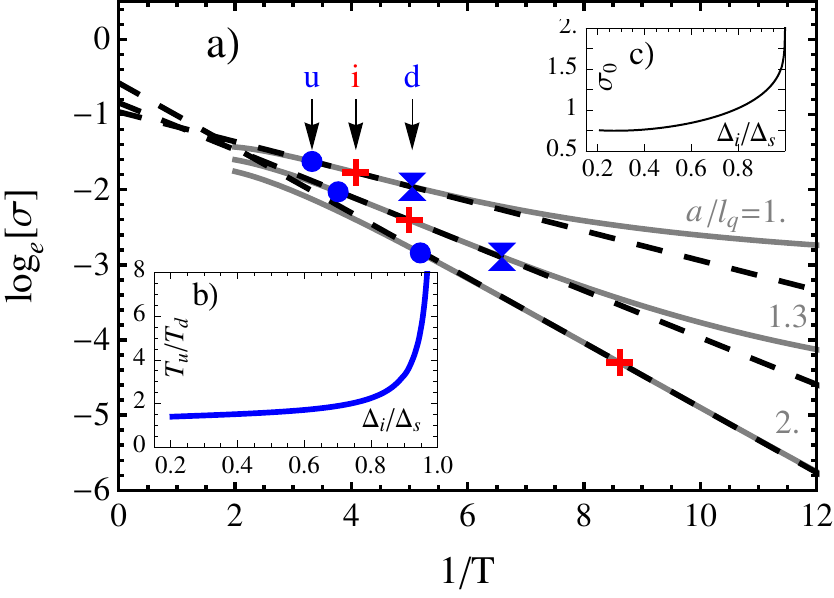}
\caption{\label{fig:sig+range}
a) Log$_e\, \sigma$ vs $1/T$ for different values of $a/\ell_q$  ($a$
is the typical saddle-point width). The estimated activation gap 
$\Delta_i$  (in units of  the saddle-point gap $\Delta_s$)
is twice the maximum value of the slope (dashed line) drawn at the inflection points
at temperatures $T_i$ (crosses). 
$\Delta_i/\Delta_s = 0.4$, $0.63$ and $0.86$ for  the illustrative values
of $a/\ell_q$.
 $\log_e \sigma$ appears linear in
$1/T$ between $T_u$ (disks) and $T_d$ (hourglasses). At $T_u$ and $T_d$, 
the slope is 95\% of its value at $T_i$. The ratio $T_u/T_d$ is shown
in b) as a function of $\Delta_i/\Delta_s$.    
c) The prefactor of the
activated behavior, estimated from the extrapolation of the linear
region to $1/T=0$, as a function of $\Delta_i$. 
}

\end{figure}

Excitations follow the classical trajectories except in the neighborhood
of a saddle point. Assuming fast equilibration in the localized regions \cite{Polyakov95,note_on_equilibration},
the conductance of the saddle point for the transport of
QPs (taken here to be moving in the $x$ direction) is given by
\begin{equation}
\sigma =  \frac{(qe)^2}{h} \frac{1}{kT}  \int_{0}^{\Delta_s} dE\;  {\cal T} (E-E_s) e^{- E /
kT},
\label{eq:current}
\end{equation}
where $E_s$ is the height of the saddle point,  ${\cal T}(E-E_s)$  is the
probability that a QP with energy $E$ is transmitted across the
saddle point and $\Delta_s$ is the energy required to excite a QP or QH at a saddle point.
When $ l_q \ll a_x $, we have ${\cal T} = 0$ or 1,
according to whether $E$ is smaller or larger than $E_s$, and we recover the
result of \cite{Polyakov95}.

The tunneling probability, $\cal T$, can
be computed  exactly in the special case of noninteracting particles
when the potential energy near the saddle  is $W=E_s -U_x x^2 + U_y y^2$ 
\cite{Fertig_Halperin87}. Provided that 
$U_{x,y}/m\omega_c^2 \approx ( \ell_q/a_{x,y})^2/2 < 1$,
$ 
{\cal T}(E-E_s) = 1/ (1 + e^{- \pi (E-E_s)/(\ell_q^2 \sqrt{U_x U_y} )} ) ,
$
and the exponent in the denominator is given correctly by the WKB
approximation. Here $\omega_c$ is the cyclotron frequency.
Assuming that 
this form for $\cal{T}$ is valid for arbitrary saddle-point potentials, we have 
computed the exponential factor within the WKB approximation and hence the average of
$\log_e \sigma$ for different (symmetric) distributions of saddle-point heights and for
different potentials. The results are not sensitive to the exact 
form assumed for the potential or the distribution of saddle-point
heights and widths, provided these are symmetrically distributed about
their mean values. The principal role of interactions is to set the energy
scale, $\Delta_s$. We ignore additional interaction effects arising 
from the non-Fermionic nature of the excitations which are thought 
to be  small \cite{MatthewsCooper09}.

Figure \ref{fig:sig+range}a shows
$\log_e \sigma$ as a function of $1/T$ for different values of the
average width, $a$. The saddle points are described by 
$W = E_s -U_x x^2 + U_y y^2$, with $\sqrt{U_xU_y}=\Delta_s/2a^2$, and
$E_s$ distributed evenly between $0$ and $\Delta_s$.  
Approximately linear dependence of the form $\log_e \sigma \sim
-\Delta_i/2T$ is observed
between $T_d$ and $T_u$, which we take to be the temperatures at which the 
gradient is within 5\% of its maximum
value.  $\Delta_i$ is defined as
twice the slope of the tangent drawn at the inflection point at temperature
$T_i$.
For $T<T_d$, the response is dominated by
activation energies lower than $\Delta_i$ but
which depend on temperature. 
For $T>T_u$, $\sigma$ is thermally activated but less
than an Arrhenius law predicts,
because of the upper limit in (\ref{eq:current}).
(Excitations with
energy above $\Delta_s$ are minority carriers localized 
in the adjacent puddle and do not contribute to transport).
An estimate of the ratio $T_u/T_d$ is shown in Fig \ref{fig:sig+range}b as a function
of $\Delta_i/\Delta_s$.
Except when $\Delta_i \gtrsim 0.85 \Delta_s$, the ratio
is around two or less. This is in line with experimental observations
\cite{Willett88,Du93,MorfdA03,Dean08}, 
which also find $\log_e \sigma \sim -\Delta_i/2T$ over a temperature range
with $T_u/T_d \lesssim 2$.

Figure \ref{fig:sig+range}c shows
the prefactor of the apparent activated behavior as a function of
$\Delta_i/\Delta_s$. 
For $\Delta_i \lesssim 0.85 \Delta_s$, the prefactor is between 0.8 and 1.25 
in units of $(qe)^2/h$  and only approaches the value two, predicted
in \cite{Polyakov95}, when the effects of tunneling are small ($\Delta_i/\Delta_s \gtrsim 0.95$).
The  rapid drop of $\sigma_0$,  as  $\Delta_i/\Delta_s$
is reduced from one, explains why 
$\sigma_0$ has been reported to lie between 0.8 and 1.1 \cite{Clarketal88}
with only a few  datapoints (mostly at $\nu=4/3$ 
close to a spin transition) falling below this region \cite{Katayama94}.
However, the use of the prefactor obtained from observation
is not easy, as it involves exponentiation of a value obtained by extrapolating to the 
limit $1/T \rightarrow 0$,
and may not be a reliable measure of the tunneling effect.

\begin{figure}[t]
\includegraphics[width=3in]{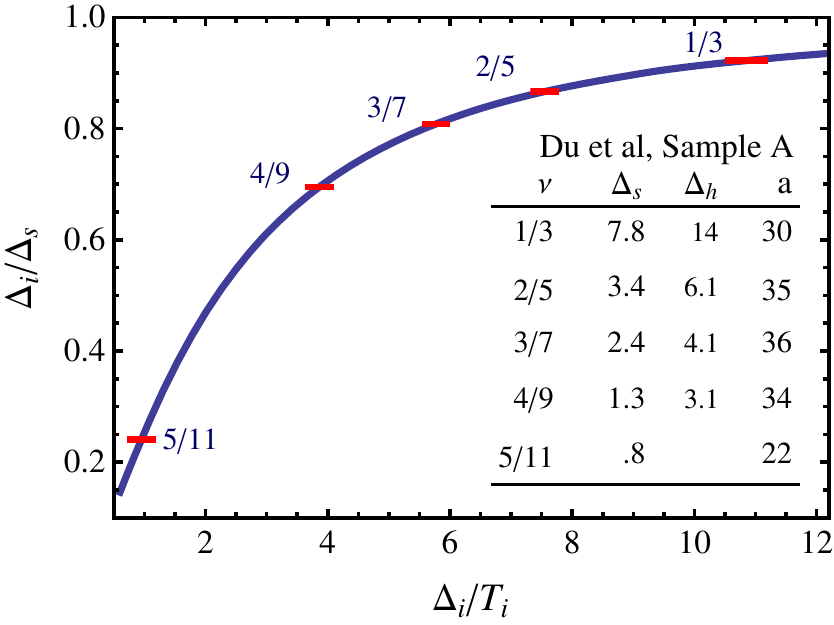}
\caption{\label{fig:FacvsInfl}
The ratio of the apparent gap to the saddle-point gap, $\Delta_i/\Delta_s$, as a function of $\Delta_i/T_i$. 
Points are placed on the curve at the values of $\Delta_i/T_i$ 
for sample A of \cite{Du93}.  This allows us to estimate
$\Delta_i/\Delta_s$ at the corresponding values of
$\nu$. Results in the table  are for $\Delta_s$  (in Kelvin)
together with the gap predicted for
homogenous systems, $\Delta_h$, and
the saddle point width at the average saddle-point, $a$ (in nm).}
\end{figure}

The extent of thermally assisted tunneling in actual samples 
can be estimated from $\Delta_i/T_i$. While the effect is 
controlled in our model by the ratio  $a/\ell_q$,  
the ratio $\Delta_i/T_i$ is directly accessible experimentally.
We have estimated  
$\Delta_i/T_i$ for sample A of \cite{Du93} (which has $d=80$nm) both
by looking for the 
inflection point and from the values for $T_u$ and $T_d$ 
(our model gives $1/T_i$ at around $0.6/T_u + 0.4/T_d$).
From the dependence of $\Delta_i/\Delta_s$ as a function of
$\Delta_i/T_i$, which is shown in Fig \ref{fig:FacvsInfl}, we obtain the estimates of
$\Delta_s$  given in the inset
table. We also include estimates for the average saddle-point width, $a$,
and the gap for ideal homogeneous
systems, $\Delta_h$ \cite{Morf_NdA_SdS_02,MorfdA03}.

Two striking features of the estimates in Fig \ref{fig:FacvsInfl} are
that the typical saddle-point widths $a$ do not vary much between
filling fractions $1/3$ and $4/9$ for the given sample and 
that the saddle-point gaps, $\Delta_s$, are consistently a factor around
two less than the values predicted for homogeneous systems,
$\Delta_h$. Given that the width of the
incompressible strips, $b$, is not expected  to vary significantly for filling
fractions in the hierarchy, we should also expect $a$ to be approximately constant.
$b$ is determined by the electrostatic potential difference, $\delta \phi$,  
between the two sides of the 
incompressible region, set up by the charge redistribution needed
to pin the charge density to its incompressible value.
If the electrostatic energy gain from transferring a QP across the region
exceeds the gap energy, QPs and QHs will be nucleated on both sides. The
width of the strip is fixed by the condition $e\delta \phi = \Delta_s / q$.
$\Delta_s/q$ is
the discontinuity in chemical potential for electrons, which is
expected to be nearly constant (in units of $e^2/\epsilon  \ell_0$) for 
hierarchy states \cite{HLR}. This leaves only a weak
dependence of $\delta \phi$, and therefore $b$, on 
$\sqrt{B}$ for states at $\nu$ as $\nu \rightarrow 1/2$, consistent
with what we find. 
In Figs \ref{fig:sig+range} and \ref{fig:FacvsInfl}, we only show results
for $\Delta_i/\Delta_s \gtrsim 0.2$ corresponding to
$(a/\ell_q)\gtrsim 0.7$. 
For small $a$, the neglect of LL mixing in the WKB calculation
is no longer justifiable, and there may also be
direct tunnelling between QH-rich and QP-rich regions. Enhanced tunnelling would appear as a
reduced value for $a$ and may explain its low value at $\nu=5/11$.

The microscopic gaps, $\Delta_s$, computed using our model from the
data of \cite{Du93} are around 55\% of  the values, $\Delta_h$, predicted using exact
diagonalizations of systems of finite numbers of particles at $\nu
=1/3$, $2/5$, $3/7$, and 40\% at $4/9$. Although the calculations
take account of the nonzero width of the quantum wells and
(perturbatively) of LL mixing, they are all for homogeneous
systems. Residual short-range scatterers, omitted from our model, may
lead to a reduction of the mobility gap \cite{wan05,Murthy09}. Furthermore,
the presence of metallic screening regions makes it unlikely that 
$\Delta_h$ would be the correct value for
$\Delta_s$.  First, $\Delta_h$ is the energy to create  a QP or QH pair
at infinite separation, while 
the excitations can be at most $O(a)$ apart across a saddle.
Second, the interaction
between particles, responsible for the gap in the
incompressible regions in fractional quantum Hall systems, will be
reduced by the screening of the nearby metallic regions. Although
this screening will affect the form of the interaction, its principal
effect in the lowest LL 
will be to reduce the overall strength of the
interaction. A reduction of the gap by a factor which is roughly 
constant is reasonable. 

We have analyzed the even-denominator states in the second LL using
data taken at zero tilt angle  on the 
sample discussed in Chapter 5 of \cite{Dean09}. This has  $n=1.6\times 10^{11}$cm$^{-2}$ and $d=160$nm.
We have extracted $\Delta_i/T_i$ from the Arrhenius plots and
estimated the saddle-point gaps, $\Delta_s$
and (using $l_q=2l_0$, $q=1/4$)  widths, $a$:
$$
\begin{tabular}{c|clllll} \hline
 $\nu$ & $\Delta_i/T_i$ & $\Delta_i$[K] & $\Delta_s$[K] & $\Delta_h$[K] & $a/l_q$ \\ \hline 
5/2 & 5.4 & 0.32 & 0.40 & 1.0 &  1.7 \\
7/2 & 1.0 & 0.035 & 0.14 & 0.85 & 0.84 \\
\hline
\end{tabular}
$$
The $\Delta_h$ have been computed for the spin-polarized
ground states in a quantum well with the same density and well width (40nm)
using exact
diagonalization for small systems and accounting for LL mixing using the RPA method of \cite{MorfdA03}.
For $\nu=5/2$, $\Delta_s/\Delta_h \approx 0.4$,
which is the same as that found for the state at $\nu=4/9$
for the sample of \cite{Du93} (see Fig. \ref{fig:FacvsInfl}).
For $\nu=7/2$, the strong tunneling across the saddle point, $a/l_q
\lesssim 1$, may explain the small $\Delta_s/\Delta_h$.
We note that, according to our model, the quality of this sample is
in part due to its large setback distance $d=160$nm.
A large value of $d$ will reduce the gradients of the impurity
potential and hence lead to increased values for $a$. According to our model,
even a small reduction in $a$ would mean the loss of activated
behavior at $\nu=7/2$ in this sample.

Data for the integer quantum Hall effect (IQH)  can also be accounted for
on the basis of thermally assisted tunnelling. However, the 
IQH is more complicated, because 
the charging energy of a localized region $\sim (qe)^2/\epsilon d$
which in the fractional quantum Hall case is much smaller than $\Delta_s$ and which we have neglected, 
is larger and is likely to be important. Nevertheless, our analysis of the data of \cite{Clarketal88}
shows that for the sample G160 $\Delta_i/T_i$ varies between nine at $\nu =2$ ($\Delta_i/\Delta_s \approx 0.9$) 
and four ($\Delta_i/\Delta_s \approx 0.65$) at $\nu = 14$. For these filling fractions, $T_u/T_d$ reduces from
just over 2 ($\nu=2$) to around 1.5 ($\nu=14$) consistent with the prediction of our model 
[Fig. \ref{fig:sig+range}(b)].

Our adaptation of the model of \cite{Polyakov95}, which takes
account of the nucleation of carriers as a result of long-range potential fluctuations and
of tunneling through saddle points, accounts for
experimental data taken on fractional quantum Hall
samples.  The dissipative transport is controlled by the energy to create excitations
close to a saddle point, $\Delta_s$.
This is smaller than the gap in an ideal homogeneous system, $\Delta_h$,
because of residual short-range scatterers and because of 
the changed nature of the interaction in the presence of nearby
metallic screening regions. 
The gap extracted from Arrhenius plots,
$\Delta_i$, is reduced from $\Delta_s$ by tunneling effects by an
amount that can be estimated via the ratio $\Delta_i/T_i$ (see Fig. \ref{fig:FacvsInfl}).
For a given filling fraction,
the largest values of $\Delta_i$ will be observed
in samples with small gradients of
the disorder potential acting on the 2DEG. This will act to increase
typical saddle-point widths, $a$, and will occur
in samples with large setback
distances, $d$, and with the most even
distribution of the ionized donors (thought to be enhanced by thermal
cycling \cite{Dean09}), both of which lead to larger values of $a$.

\begin{acknowledgments}
We thank R.N. Bhatt for helpful remarks and G. Gervais for help with \cite{Dean08,Dean09}.
The work was supported in part by the NSF under Grant 
DMR-0906475. 
\end{acknowledgments}
\bibliographystyle{apsrev4-1}
\bibliography{model_disorder}

\begin{thebibliography}{22}
\bibitem [{not()}]{note_on_equilibration}%
  \BibitemOpen
  \href@noop {} {}\bibinfo {note} {With reduced drift velocities (due to
  screening) and Coulomb scattering off particles in the compressible regions
  as the dominant scattering mechanism, the condition for equilibration is more
  easily satisfied than in \cite{Polyakov95} (where phonon scattering is the
  dominant mechanism).}\BibitemShut {Stop}%
\end{thebibliography}
\end{document}